\documentstyle{article}
\begin{document}
\vspace{-2cm}
\title{\sc Leading Particle Effect in the  $J/\psi$ Elasticity Distribution} 
\author{F.O. Dur\~aes$^{1,2}$\thanks{e-mail: 
fduraes@if.usp.br}, \ F.S.
Navarra$^{2,3}$\thanks{e-mail: navarra@if.usp.br} \ and \ G.
Wilk$^{3}$\thanks{e-mail: Grzegorz.Wilk@fuw.edu.pl} \\[0.1cm]
{\it $^1$Instituto Superior T\'ecnico, Universidade T\'ecnica de
Lisboa}\\
{\it Av. Rovisco Pais 1, 1-1096 Lisboa Codex, Portugal}\\[0.1cm]
{\it $^2$Instituto de F\'{\i}sica, Universidade de S\~{a}o Paulo}\\
{\it C.P. 66318,  05389-970 S\~{a}o Paulo, SP, Brazil} \\[0.1cm]
{\it$^3$Soltan Institute for Nuclear Studies, 
Nuclear Theory Department}\\
{\it ul. Ho\.za 69, \ 00-681 Warsaw, Poland}}
\maketitle
\vspace{1cm}
\begin{abstract}

Recent DESY-HERA data on $J/\Psi$ elasticity distribution show
that it emerges mostly as a fast particle. Interpreting photoproduction
as a collision between a pre-formed charmed hadron and the proton, the
outcoming  $J/\Psi$ is a leading particle of the collision. We 
analyse these data using a model formulated to describe  energy flow in
hadron-hadron reactions. The measured $J/\Psi$ spectrum can be successfully
described  in terms of this model. We conclude that the observed transparency 
of the charmed hadron-proton collisions arises because of the particularly
small gluonic content of the initial $c - \overline c$ state.\\

PACS number(s): 13.85.Qk, 11.55.Jy
\end{abstract}

\vspace{1cm}

The leading particle effect is one of the most interesting features of 
multiparticle production in hadron-hadron collisions \cite{ehs}. In these
reactions the valence quarks of the projectile emerge from the collision 
carrying the initial state quantum numbers. All produced particles come 
essentially from the gluons and quark-anti-quark pairs already pre-existing 
in the projectile and target, or radiated during the collision. This 
qualitative picture takes different implementations in the many existing
multiparticle production models. In one of them, the Interacting Gluon
Model (IGM) \cite{IGM,IGMC,IGMD,IGMH}, the produced particles (and 
consequently the energy released in the secondaries and lost by the 
projectiles) come almost entirely from the pre-existing gluons in the
incoming hadrons. This conjecture may be directly  tested using a 
high energy, nearly gluonless hadronic projectile. In this case, according
to the IGM, inspite of the high energy involved, the production of
secondaries would be suppressed (in comparison to the production observed
in ordinary hadron induced reactions) and the energy would be mostly carried
away by the projectile leading particle (LP) which would then be observed 
with a hard $x_F$ spectrum. This type of gluonless projectile is available
in   $J/\Psi$ photoproduction, where  the photon can be understood as a 
virtual 
$c- \overline c$ pair which reacts with the proton and turns into the
finally observed  $J/\Psi$. There are low energy data taken by the FTPS
Collaboration \cite{ftps} and 
very recently high energy data became available at HERA \cite{JPSIDATA}.

It has been shown by a number of authors that
the simple vector meson dominance mechanism (VDM), in which the photon is
converted directly to a  $J/\Psi$ before the interaction with the proton, 
does not describe many aspects of data. However, as it was shown in ref. 
\cite{HK}, the charm anti-charm pair can be understood as a 
supperposition of many hadronic states: $J/\Psi$, $\Psi^{'}$, ... . Whereas
relevant for the calculation of the cross sections, the inclusion of all the
other charmonium states does not change the fact that the projectile, 
whatever it may be, is a hadron with low gluonic content. For the study of 
the outgoing $J/\Psi$ momentum spectrum, this is the most important 
information, which allows us to test the picture proposed by the IGM. 

At lower energies we can compare the momentum spectrum of 
the  $J/\Psi$  
measured  in $\gamma\,p\,\rightarrow\,J/\Psi\,X$ collisions \cite{ftps} 
with the leading meson ($\pi$ and $K$) momentum spectra measured 
\cite{ehs} in $\pi\,p\,\rightarrow\,\pi\,X$ 
and $K\,p\,\rightarrow\,K\,X$ 
reactions at the same c.m.s.  energy. One observes that the charmed  
leading particles are 
much harder. This comparison is however not completely meaningful because
the  $J/\Psi$ data contain a diffractive component which was subtracted in 
the hadronic data. 

At HERA a  $J/\Psi$ $z$  ($z = E_{J/\Psi}/E_{\gamma}$) 
spectrum presumably free from 
the diffractive component was presented. Although the energy is different 
a comparison with the leading particle spectra measured in 
hadronic reactions is still
meaningful because the LP spectra have a relatively weak energy dependence.
The  $J/\Psi$ momentum spectrum is clearly harder, according to what we 
expect in the IGM. The $\Upsilon$ $z$ spectra, which will be eventually 
measured at HERA in a near future, will be even harder.

Although there are already some  
PQCD calculations
of $J/\Psi$ photoproduction \cite{PQCD} the large $z$ values relevant 
here ($0.5 \leq z =E_{J/\Psi}/E_{\gamma} \leq0.9$), for 
which PQCD is not always applicable \cite{PQCD}, justify, in our
opinion, the more phenomenological examination of these spectra. In  
$J/\Psi$ hadroproduction perturbative calculations fail at large $x_F$, 
where non-perturbative effects become stronger. Here, these effects 
may also play some role in the large $z$ region.

We present now our quantitative calculations, which illustrate the
qualitative discussion made above. The IGM is a model which
has been developed primarily to analyse energy-momentum spectra of
leading particles \cite{IGM,IGMC,IGMD,IGMH}. 
This work continues therefore the application of the
IGM to photon initiated reactions presented in \cite{IGMH} putting
them  on an equal footing with the
hadronic processes studied before \cite{IGM,IGMC} (including
diffractive dissociation ones \cite{IGMD}).\\ 

In Fig. 1 we show schematically the IGM picture of a photon-proton
collision. According to it, during the interaction the photon is
converted into a hadronic state which interacts with
the incoming proton. This hadronic state contains the  
$c\,-\,\overline c$ and some gluons and we call it simply ``hadron''.
The hadron-proton interaction follows 
then the usual 
IGM picture, namely: the valence quarks fly through essentially
undisturbed whereas the gluonic clouds  of both projectiles interact
strongly with each other \footnote{By gluonic clouds we understand a
sort of ``effective gluonsn'' which include also their fluctuations seen
as $\bar{q}q$ sea  pairs.}. In the course of interaction the hadronic
state looses fraction $x$ of its original momentum and gets excited
forming what we call a {\it leading jet} (LJ) which carries fraction
$z = 1 -x$ of the initial momentum. The proton looses fraction $y$ of
its momentum forming another leading jet. In the IGM we consider two
possible types of $\gamma + p$ interactions: non-diffractive and
diffractive. In each of these reactions the $J/\Psi$ can come either 
from the fragmentation of
the mesonic leading jet or from the hadronization of the central
gluonic fireball. In this work  we shall concentrate only on the data
taken at  HERA \cite{JPSIDATA} for  $p^2_T\ge1$ GeV$^2$ and
$0.5\le z \le 0.9$. In this case, it is enough to consider  
the single non-diffractively $J/\Psi$ produced from the fragmentation
of the photonic leading jet, cf. Fig. 1. All other contributions can
be safely neglected here. 

In order to calculate this spectrum we start 
(cf., \cite{IGMD,IGMH} for details)
with the function $\chi(x,y)$, which describes the probability to
form a central gluonic fireball (CF) carrying momentum fractions $x$
and $y$ of the two colliding projectiles: 
\begin{eqnarray}
\chi(x,y) &=& \frac{\chi_0}{2\pi\sqrt{D_{xy}}}\cdot \nonumber\\
&&\cdot \exp \left\{ - \frac{1}{2D_{xy}}\,\left[
  \langle y^2\rangle (x - \langle x\rangle )^2 +
  \langle x^2\rangle (y - \langle y\rangle )^2 -
  2\langle xy\rangle (x - \langle x\rangle )(y - \langle y\rangle )
  \right] \right\}, \label{eq:CHI}
\end{eqnarray}
where
\begin{equation}
D_{xy} = \langle x^2\rangle \langle y^2\rangle - 
           \langle xy\rangle ^2 \qquad {\rm and}\qquad
\langle x^ny^m\rangle = \int_0^1\! dx\,x^n\, 
\int_0^1\! dy\, y^m\, \omega (x,y). \label{eq:defMOM}
\end{equation}
Here $\chi_0$ denotes the normalization factor provided by the
requirement that $ 
 \int_0^1\!dx\, \int_0^1\! dy\, \chi(x,y) \theta(xy - K_{min}^2) = 1$
with $K_{min} = \frac{m_0}{\sqrt{s}}$ being the minimal inelasticity
defined by the mass $m_0$ of the lightest possible central fireball CF
(represented by the blob in Fig. 1). The dynamical input of the IGM is
contained in the, so called, spectral function $\omega(x,y)$ given by
\begin{equation}
\omega(x,y)\, =\, \frac{\sigma_{gg}(xyW^2)}{\sigma(W)}
   \, G(x)\, G(y)\, \theta\left(xy - K^2_{min}\right),
   \label{eq:OMEGA}
\end{equation}
where $G(x)$ and $G(y)$ denote the effective number of gluons in the
charmed hadron and in the proton, respectively. They are approximated by the
respective gluonic structure functions. $W$ is the photon-proton c.m. 
energy. $\sigma_{gg}$ is the gluon-gluon cross section, which is
computed over a wide range of scales given by $xyW^2$. Whenever the
scale is larger than $2.3$ GeV lowest order perturbative QCD formulas
are used. Otherwise a parametrization which represents
non-perturbative physics is employed \cite{IGM}. At the energies $W$
considered here the bulk of the interaction happens in the
non-perturbative domain. $\sigma$ is the relevant meson-proton cross
section.\\

The final momentum spectrum of the produced $J/\Psi$ is then given in
terms of $\chi(x,y)$ as follows:
\begin{eqnarray}
F (z) &=& \int^1_0\! dx\int^1_0\! dy\,\, \chi(x,y)\,
             \delta(z-1+x)\, \theta\left(xy - \frac{m_0^2}{W^2}\right) 
             \, \theta\left[y -
             \frac{(M_{J/\Psi}+m_0)^2}{W^2}\right] \label{eq:PSILEAD}\\
          &=& \int^1_{y_{min}}\!\! dy\, \chi(x=1-z;y)\nonumber
\end{eqnarray}
where
\begin{equation}
y_{min} = Max\left[\frac{m_0^2}{(1-z)W^2},\, 
   \frac{(M_{J/\Psi}+m_0)^2}{W^2}\right]  \label{eq:limits}
\end{equation}
and $z = \frac{E_{J/\Psi}}{E_{\gamma}}$ is the $J/\Psi$ 
energy fraction (which in the IGM, where all masses have been
consistently neglected coincides with the momentum fraction). Because
we are dealing here with the leading particle spectra, we have to
introduce the additional kinematical constraint, $y >
\frac{(M_{J/\Psi}+m_0)^2}{W^2}$, which ensures that the  mass $M_X$
($M_X= \sqrt{y} W$, see Fig. 1)  is large enough to produce both the
measured $J/\Psi$ particle of mass $M_{J/\Psi}$ and the minimal CF of
mass $m_0$, as demanded by the IGM. (In fact, in our case we have to
replace $M_{J/\Psi}$ by $M_T = \sqrt{M^2_{J/\Psi} + p^2_T},~p_T^2 = 
1$ GeV$^2$, to account for the minumum transverse momentum present at
our data points).\\ 

Whenever possible we keep all parameters the same as in the
previous applications of the IGM \cite{IGM} - \cite{IGMH}. However,
the inelastic $ (c-\overline c)\,-\,p$ scattering cross section and the 
number of gluons in the charmonium state are different from the 
corresponding quantities encountered so far. 

The charmonium-hadron cross section, which we shall approximate by
$\sigma^{inel}_{J/\psi-p}$, 
has been subject of intensive research in the context of nuclear
physics and signatures of quark gluon plasma. Calculations seem to
converge to $\sigma^{inel}_{J/\psi-p} \simeq 6-9$ mb \cite{satz}.
As for the distribution $G(x)$, which we may call $G^{J/\psi}(x)$, 
we shall assume that it has
the same shape as in other mesons, i.e., $ G^{J/\psi}(x) =
G^{\rho^{0}}(x) = G^{\pi}(x)$ and use, for the latter, the SMRS
parametrization \cite{sutton}. The specific shape chosen for these
distributions does not affect much the results. Their normalization
$p^h = \int_0^1\! dx\,x\,G^h(x)$, i.e., the amount of momentum
allocated to the gluonic component, plays, however, a crucial role
here and we should try to estimate it somehow. It is known that in a
nucleon or in a light meson $ p^h  \simeq 0.5 $, i.e., gluons carry
half of the momentum of the hadron. The charmonium, however, is a
non-relativistic system and almost all its mass comes from the quark
masses. The gluonic field, responsible for a weak binding, carries
only a small fraction of the energy (and momentum) of the bound
state. We expect therefore the normalization factor $p^{J/\psi}$ of
$G^{J/\psi}(x)$ to be of the  order of the energy stored in the field
divided by the mass of the state. In the case of a $J/\psi$ we have:  
\begin{eqnarray}
p^{J/\psi} \,& = & \, \frac{M_{J/\psi}-2\,m_c}{M_{J/\psi}}\, 
\simeq \, 0.033, \label{eq:p-psi}
\end{eqnarray}
where $m_c$ has been taken $1.5$ GeV and $M_{J/\psi}=3.1$ GeV.
In the  IGM those two parameters are in fact entering only as a combination
$p^{J/\Psi}/\sigma^{inel}_{J/\Psi-p}$. Because out of those two
quantities relevant here only the $p^{J/\Psi}$ is so far completely
unknown, we shall, in what follows, take for definiteness
$\sigma^{inel}_{J/\Psi-p} = 9$ mb and leave $p^{J/\Psi}$ to be a free
parameter.

In Fig. 2 we compare our results ($F(z)$) with the experimental
data. As it can be seen the agreement is very good and it was
obtained with essentially only one new (but heavily constrained)
parameter equal to the ratio of the amount of momentum carried  by 
gluons in the $c\,\overline c$ state
and its inelastic cross section with the proton.
The observed flatteness comes (once $\sigma^{inel}_{J/\Psi-p}$ is
fixed) entirely from the small value of the $p^{J/\Psi}$ parameter,
what can be seen if we compare our best result 
($p^{J/\Psi}\,=\,0.033$) shown in 
the full line, with
results for other values of $p^{J/\Psi}$. The dashed line represents
the choice  $p^{J/\Psi}\,=\,0.066$ and the dotted line corresponds to 
$p^{J/\Psi}\,=\,0.016$. Notice the high sensitivity
of the results to the changes of the parameter $p^{J/\Psi}$.

To summarize: assuming that the photon may be represented by a 
hadronic state containing a charm anti-charm pair,
we have successfully described the leading spectra of  
photoproduced $J/\Psi$'s in terms of the IGM. At the same time
we have demonstrated that (again: within the IGM scheme) they
depend crucially on the amount of momentum carried by gluons in the
charmed hadron, $p^{J/\Psi}$ (provided its cross section 
$\sigma^{inel}_{J/\Psi-p}$ with the nucleon is known from 
somewhere else, otherwise they depend of the ratio of these
two parameters). It means that the knowledge of the leading 
spectra and the inelastic cross section should 
allow us to estimate the amount of the gluonic momenta in the 
projectile. For high energies (as those encountered here) this 
result is universal and  insensitive to the mass of the 
projectile under consideration.

\vspace{0.5cm}
\underline{Acknowledgements}: This work has been supported by CNPq, CAPES
and FAPESP under contract number 93/2463-2.

\noindent
{\bf Figure Captions}\\
\begin{itemize}

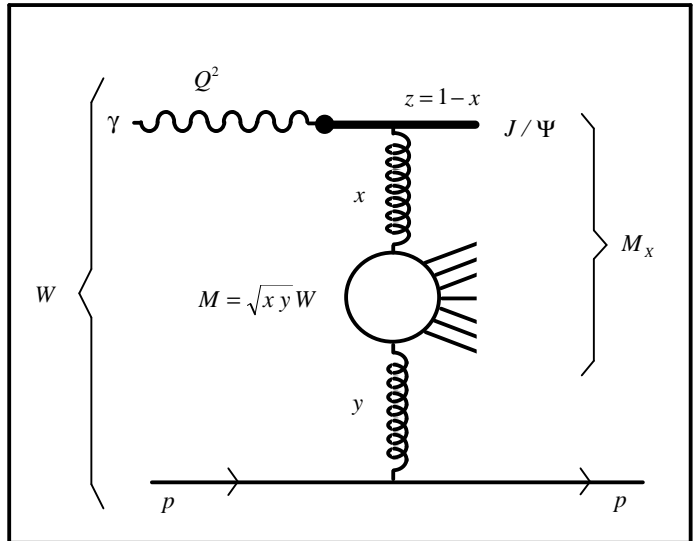
\item[{\bf Fig. 1}] IGM description of a photon-proton scattering
with  $J/\psi$ production.

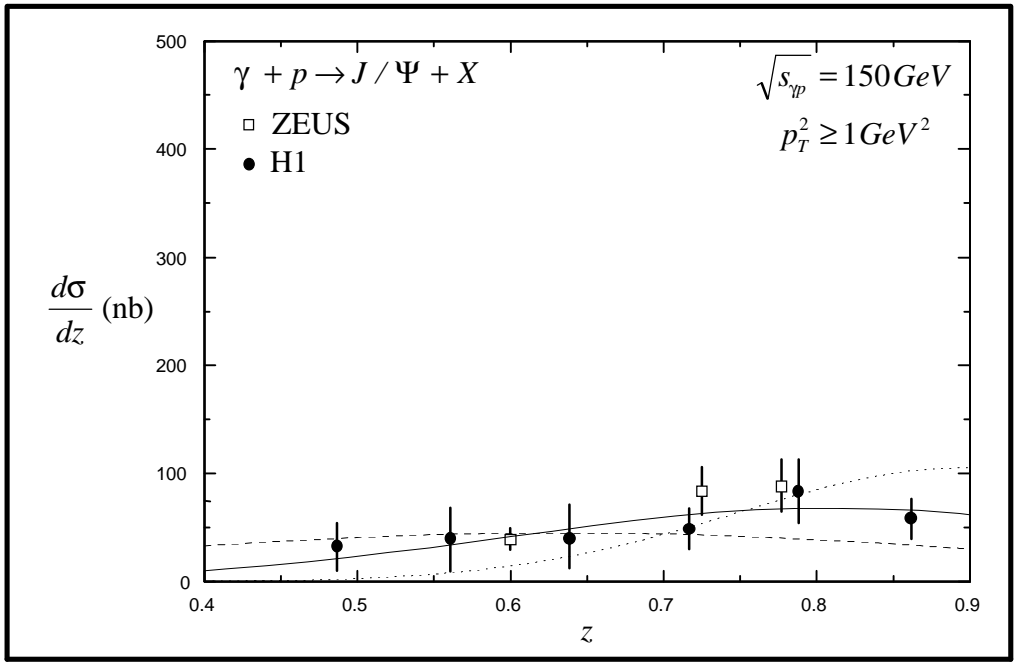
\item[{\bf Fig. 2}] Comparison of the IGM distribution $F(z)$
with data of ref. \cite{JPSIDATA} with  restricted acceptance $p_T^2
\geq 1\, (GeV/c)^2$ and $0.5 \leq z \leq 0.9$ for fixed value of
$\sigma^{inel}_{J/\Psi-p}=9$ mb and for three different values of
$p^{J/\Psi}$: $0.066$ (dashed line),  $0.033$ (solid line) and  
$0.016$ (dotted line).
\end{itemize}


\begin{thebibliography}{99}

\bibitem{ehs} EHS/NA22 Collab., N.H. Agababyan et al., {\sl Z. Phys.} 
               {\bf C75} (1996) 229. 

\bibitem{IGM} F.O.Dur\~aes, F.S.Navarra and G.Wilk, {\sl Phys.
                Rev.} {\bf D47} (1993) 3049 and references therein.

\bibitem{IGMC} F.O.Dur\~aes, F.S.Navarra, C.A.A. Nunes and G.Wilk, 
                {\sl Phys. Rev.} {\bf D53} (1996) 6136. 

\bibitem{IGMD} F.O.Dur\~aes, F.S.Navarra and G.Wilk, {\sl Phys.
                Rev.} {\bf D55} (1997) 2708.
                
\bibitem{IGMH} F.O.Dur\~aes, F.S.Navarra and G.Wilk, {\sl Phys. Rev.}
               {\bf D56} (1997) R2499.


\bibitem{ftps} B.H. Denby et al. {\sl Phys. Rev. Lett.} 
               {\bf 52} (1984) 795. 


\bibitem{JPSIDATA} S.Aid et al., (H1 Collab.), {\sl Nucl. Phys.}  
                   {\bf B472} (1996) 3; M.Derrick et al. (ZEUS 
                   Collab.), {\sl Phys. Lett.} {\bf B350} (1995) 120;
                   J.Breitweg et al. (ZEUS Collab.), {\sl Z. Phys.} 
                   {\bf C76} (1997) 599.

\bibitem{HK} J.H\"ufner and B.Z.Kopeliovich. {\it $J/\Psi N$ and
             $\Psi 'N$ total cross sections from photoproduction 
             data: failure of vector dominance}, hep-ph/9712297.



\bibitem{PQCD} Cf., for example, P.Hoyer, {\sl Nucl. Phys.} {\bf A622}
               (1997) 284c and {\it Charmonium production at ELFE energies}, 
               hep-ph/9702385, and references therein; M. Kramer, {\sl Nucl. 
               Phys.} {\bf B459} (1996) 3; M. Cacciari and M. Kramer, 
               {\sl Phys. Rev. Lett.} {\bf 76} (1996) 4128.

               
\bibitem{satz} D. Kharzeev and H. Satz, {\sl Phys. Lett.} {\bf B366} 
               (1996) 316; {\bf B356} (1995) 365; {\bf B334} (1994) 155.
 
\bibitem{sutton} P.J. Sutton, A.D. Martin, R.G. Roberts and W.J. 
                 Stirling, {\sl Phys.Rev.} {\bf D45},  (1992) 2349.

\end{thebibliography}
\end{document}